\documentclass{aa}
\usepackage{txfonts}
\usepackage{graphicx}
\usepackage{color}
\usepackage{epstopdf}

\usepackage{natbib}
\bibpunct{(}{)}{;}{a}{}{,} 
\begin{document}

\title{Solar Flare X-ray Source Motion as a Response to Electron Spectral Hardening}

\author{Aidan M. O'Flannagain \inst{1}
\and Peter T. Gallagher \inst{1}
\and John C. Brown \inst{2}
\and Ryan O. Milligan \inst{3}
\and Gordon D. Holman \inst{4}}

\institute{Astrophysics Research Group, School of Physics, Trinity College Dublin, Dublin 2, Ireland
\and Astronomy and Astrophysics Group, School of Physics and Astronomy, University of Glasgow, Glasgow G12 8QQ, Scotland
\and Astrophysics Research Centre, School of Mathematics and Physics, Queen's University Belfast, Belfast BT7 1NN, N. Ireland
\and Solar Physics Laboratory, Heliophysics Science Division, NASA Goddard Space Flight Center, Greenbelt MD 20771, USA}

\abstract{Solar flare hard X-rays (HXRs) are thought to be produced by nonthermal coronal electrons stopping in the chromosphere, or remaining trapped in the corona. The collisional thick target model (CTTM) predicts that more energetic electrons penetrate to greater column depths along the flare loop. This requires that sources produced by harder power-law injection spectra should appear further down the legs or footpoints of a flareloop. Therefore, the frequently observed hardening of the injected power-law electron spectrum during flare onset should be concurrent with a descending hard X-ray source.}
{To test this implication of the CTTM by comparing its predicted HXR source locations with those derived from observations of a solar flare which exhibits a nonthermally-dominated spectrum before the peak in HXRs, known as an early impulsive event.}
{HXR images and spectra of an early impulsive C-class flare were obtained using the \textit{Ramaty High-Energy Solar Spectroscopic Imager} (RHESSI). Images were reconstructed to produce HXR source height evolutions for three energy bands. Spatially-integrated spectral analysis was performed to isolate nonthermal emission, and to determine the power-law index of the electron injection spectrum. The observed height-time evolutions were then fit with CTTM-based simulated heights for each energy, using the electron spectral indices derived from the RHESSI spectra. }
{The flare emission was found to be dominantly nonthermal above $\sim$7~keV, with emission of thermal and nonthermal X-rays likely to be simultaneously observable below that energy. The density structure required for a good match between model and observed source heights agreed with previous studies of flare loop densities.}
{The CTTM has been used to produce a descent of model HXR source heights that compares well with observations of this event. Based on this interpretation, downward motion of nonthermal sources should indeed occur in any flare where there is spectral hardening in the electron distribution during a flare. However, this would often be masked by thermal emission associated with flare plasma pre-heating. As yet, flare models that predict transfer of energy from the corona to the chromosphere by means other than a flux of nonthermal electrons do not predict this observed source descent. Therefore, flares such as this will be key in explaining this elusive energy transfer process.}

\titlerunning{Downward-moving Thick Target X-Ray Emission}
\authorrunning{O'Flannagain et al.}
\maketitle

\section{Introduction}

Solar flares are the largest explosions in the solar system, releasing energy on the order of $10^{25}$~J ($10^{32}$~erg) as radiation across the spectrum in a matter of minutes \citep[e.g.,][]{ems04,ems05}. The analysis of nonthermal X-ray emission is extremely important in explaining the process which causes such impulsive energy release. Observable properties such as nonthermal X-ray source position are expected to depend on the nature and evolution of the accelerated electron spectrum. However, nonthermal emission is frequently masked by thermal emission in the early phase of the flare, making it difficult to investigate nonthermal processes before the peak in hard X-rays (HXRs). There exist a small number of recorded events in the database of \textit{Ramaty High Energy Spectroscopic Imager} \citep[RHESSI;][]{lin02} called `early impulsive' flares, which can be identified by a delay of $\sim$30~s or less between the initial rise in soft X-ray flux and the impulsive rise in HXR flux. \citet{sui07} outline analysis of 33 such events, in which plasma preheating is minimal, and so nonthermal emission may be the primary contributor to the RHESSI spectrum even before the peak in HXRs. Due to their dominantly nonthermal spectra, early impulsive flares are essential in gaining an understanding of the behaviour of nonthermally accelerated electrons at the earliest phases of an event.
 
The standard model of solar flare HXR emission is believed to begin with a process of energy release in the corona, possibly magnetic reconnection \citep{swe69, pet64} and the acceleration of electrons towards the thick target chromosphere \citep{bro71,hud72}. Here, the electrons produce HXRs by nonthermal bremsstrahlung and collisionally heat the chromosphere, resulting in upward expansion of plasma which fills the post-flare loop - a mechanism known as chromospheric evaporation \citep{bro73, ant78, mil06a, mil06b}. However, to produce observed HXR fluxes, the collisional thick target model (CTTM) first requires a very large number of electrons in the tenuous corona, and then a challengingly large flow of electrons towards the chromosphere \citep{bro09}. The former so-called ``number problem'' \citep{bro77} is resolved by the formation of return currents \citep{kni77, col78}, which is discussed in recent work by \citet{zha06} and \citet{hol12}. However, the beam flux problem remains challenging especially for the small HXR source areas suggested in some RHESSI flare data \citep[e.g.,][]{kru11}. This has led to proposals of alternatives to the usual thick target injection geometry with acceleration or reacceleration of electrons within the chromosphere, by cascading small scale reconnection \citep{bro09}, or by the Poynthing flux of an Alfv\'en wave train \citep{fle08}. These models differ in the interpreted location of major particle acceleration during the early stages of the flare, and so can be tested by analysing HXRs in the corona.

Nonthermal coronal X-ray sources have previously been suggested as evidence for coronal magnetic reconnection \citep{fro71,mas94} and plasmoid-looptop reconnection \citep{mil10}. In the RHESSI era, numerous studies have been carried out on occulted flares, where the bright nonthermal footpoint emission is masked by the solar limb, allowing observations of possibly nonthermal looptop emissions which are normally outside of the dynamic range of the instrument \citep[e.g.,][]{bal02, kru07}. Coronal nonthermal emission has been shown to be temporally correlated with Type III radio bursts \citep{kru08}, further supporting the argument for the existence of a nonthermally accelerated electron population in the corona. Looptop source motion has previously been interpreted as a signature of transition from X-type to Y-type reconnection during a flare \citep{sui03}.

Early impulsive flares provide an opportunity to observe faint looptop nonthermal emission without sacrificing information on the behaviour at the footpoints during the HXR peak. This therefore allows for the detection of any source motion between the coronal looptop and chromospheric footpoints. During the rise phase of a typical flare, the flux of HXRs reach a peak and the spectral index hardens \citep{par69,ben77,fle02}. Based on the theoretical derivations of nonthermal X-ray intensity with height in the coronal acceleration scenario \citep{bro75}, this is expected to result in a descent of the location of peak nonthermal emission in the time coming up to the HXR peak. It was suggested that this downward motion of nonthermal X-ray sources was observed in the C1.1 class early impulsive flare that occurred on 28 November 2002 \citep[SOL2002-11-28T04:37,][]{sui06}. In this event, a faint looptop source appeared, split into two, and descended down both loop legs, and reached the footpoints at the time of the peak in HXRs. An in-depth analysis of this behaviour will help to test the thick target model during a phase of nonthermal emission which is rarely observed.

In this paper, observations of descending X-ray sources are modelled by taking into account the time variation in the spectral index of the electron injection spectrum. We suggest that a descent of HXR sources in the rise phase of a flare can be explained by hardening of the electron injection spectrum. In \S 2 the 28 November 2002 flare observations and analysis are presented. In \S 3 the model used to determine theoretical source positions is described, predicting the dependence of source height on spectral index and observed photon energy. In \S 4 the results of this analysis are shown, and in \S 5 interpretations are drawn based on the comparison of our theoretical models and these observations.

\section{RHESSI Observations}

\begin{figure*}
\centering
	\includegraphics[width = 18cm]{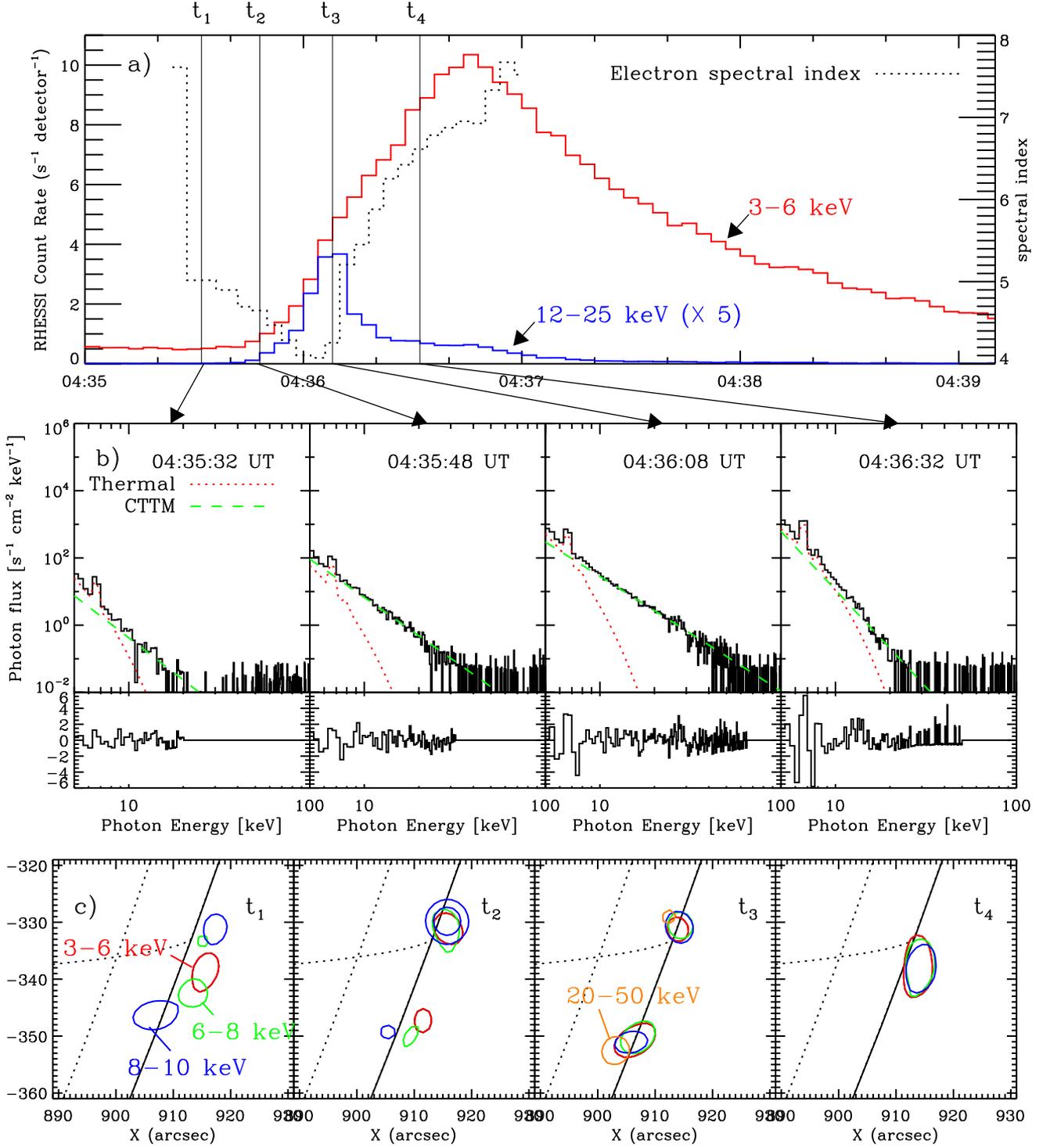}
 	\caption{(\textit{a}) X-ray Lightcurve of the flare of 28 November 2002. The 12--25~keV curve is scaled by a factor of 5 for clarity. Four times are marked, corresponding to the start times of the four images and spectra shown below. Overplotted is the electron power-law index derived from the spectral fits (dotted line), demonstrating the concurrence of maximum spectral hardness (minimum spectral index) with peak in HXR emission. (\textit{b}) Spatially integrated spectra for the times of their corresponding images using pre-flare background subtraction. Overlayed are thermal and nonthermal fits constructed using the OSPEX spectral analysis suite
. Residuals, or the difference between observed and model-based X-ray flux, normalized to the one-sigma uncertainty in the photon flux, are plotted below each spectrum. 
(\textit{c}) RHESSI image contours corresponding to energy bands of 3--6, 6--8, and 8--10~keV, with a  contour showing 20--50~keV at $\mathrm{t_{3}}$, the HXR peak. Contours represent 75\% of the peak emission of the image, with a second 50\% contour included for the 8--10~keV image at interval $\mathrm{t_{2}}$, in order to show the location of the southern source. Images are generated using the CLEAN algorithm available in the RHESSI image analysis software
. Each of the intervals used for these images are 8 seconds in duration, beginning at the time shown in (b).
}
	\label{fig:intro}
\end{figure*}

\begin{figure*}
\centering
	\includegraphics[width = 17cm]{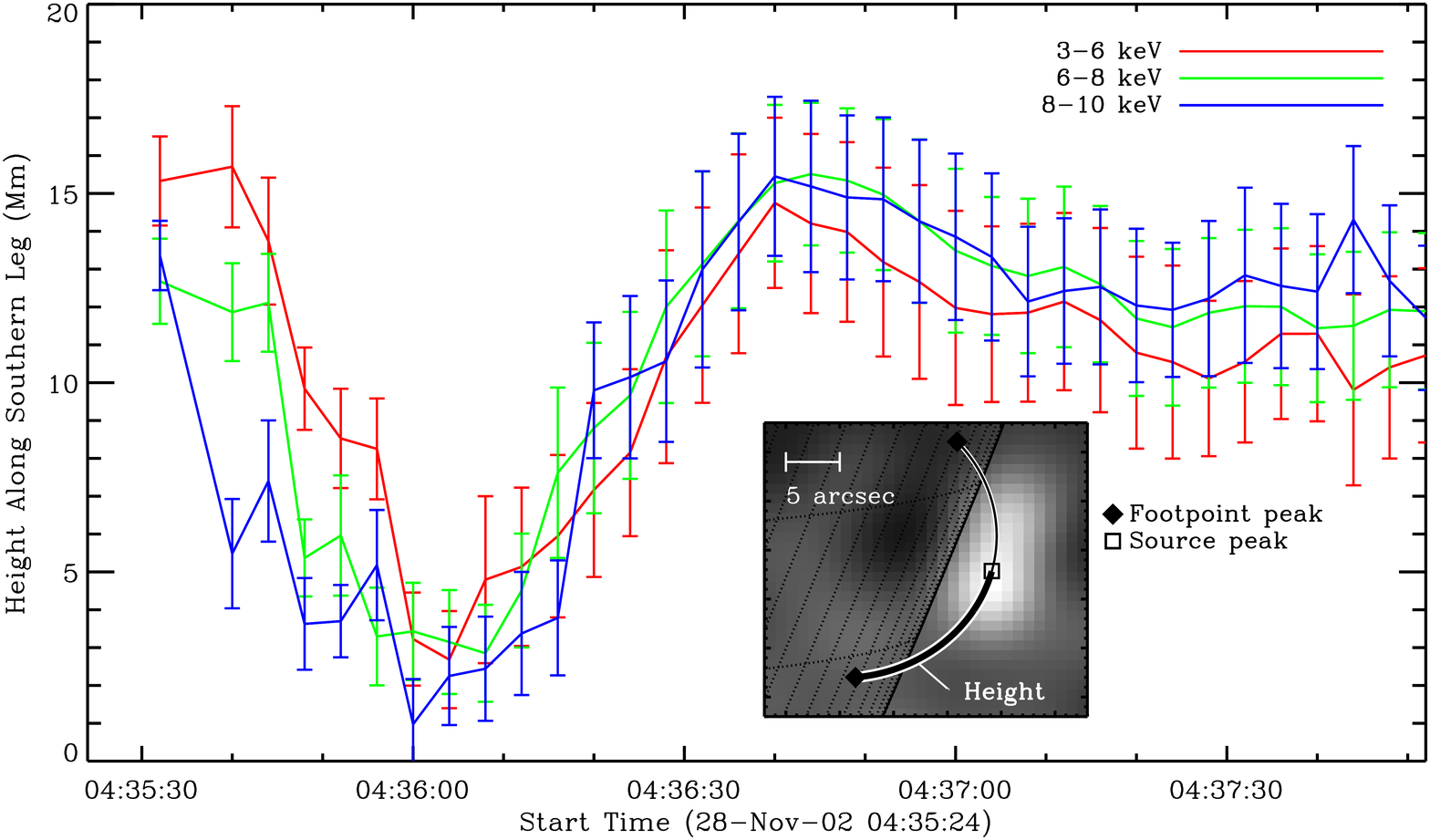}
 	\caption{ Height of X-ray source peak with time for the 3--6, 6--8 and 8--10~keV energy bands. Vertical lines respresent the 1-sigma width of the 2-dimensional Gaussian which was fit to the RHESSI source in order to determine peak location, thereby illustrating the size of the source, which is sensitive to the PSF of the instrument. Height is defined as distance in megametres from the source peak to the southern footpoint along the circle defined by the source peak position itself and both footpoints (\textit{see inset}). Footpoints are defined as the peak position of 25-50~keV emission at 04:36:08--04:36:12 UT, the peak in HXRs. The temporal spacing of the data points here does not represent the integration time of the associated RHESSI images. For all images but the first, the integration time is 8~s, while the spacing between them is 4~s, resulting in an overlap of 4~s. \textit{Inset:} An example image of 3--6 keV emission at 04:35:40--04:35:48 UT. The source, just prior to splitting into two, can be seen to the right of the image, at the assumed looptop. Overlayed on the image are locations of the peaks of gaussian fits to the current descending source (open square) and the 20--50~keV footpoints seen at the HXR peak (filled diamonds). The definition of height is visualised as the distance along the circle between the southern footpoint and the southern source.}
	\label{heighttime}
\end{figure*}

A C1.1-class solar flare was observed by RHESSI on 28 November 2002, beginning at 04:35:30 UT, with HXR emission observed for roughly 50~s (Figure 1a). The flare was located near the Sun's western limb, with unocculted footpoints. RHESSI was in attentuator state A0, meaning there were no aluminium attentuators in front of the detectors during the event. As a result, RHESSI was able to detect X-rays with energies as low as 3~keV. Throughout the event, flare emission was observed up to energies of $\sim$50~keV.

Time intervals were selected to produce as many independent images as possible without creating noise-dominated X-ray source maps of this low-count flare detection. One 16s interval was used from the start of the flare at 04:35:24 UT until 04:35:40 UT. From that point on, images were made by integrating flux over 8s, until the end of the final interval at 04:38:00 UT. In order to aid in the automated tracking of source peaks, overlapping time intervals were laid in between each of these intervals, resulting in a total of 36 images per chosen energy band. Energy bands were selected to focus on the low-energy part of the spectrum, and were set at 3--6~keV, 6--8~keV, and 8--10~keV, producing reliable imaging of source motion in all energy ranges. Images produced using higher energy bands were noise-dominated for all time intervals except during the peak in HXRs, and so were excluded from this analysis, with the exception of 25--50~keV emission at the HXR peak, which was used to estimate the location of the flare loop footpoints.

Figure 1 gives a summary of the RHESSI observations. The descent of X-ray sources down two legs of an apparent loop documented by \citet{sui06} is immediately evident upon study of RHESSI images (Figure 1c). A crucial step in modelling this behaviour was determining at what times and energies emission appeared to be nonthermal, especially within the energy range of 3--10~keV, well below the more common estimates of upper limits to the nonthermal low-energy cutoff of $\sim$20--40~keV \citep[e.g.,][]{hol03}. However, more recent work which corrects for albedo effects suggests cutoffs of less than~12keV \citep[][for recent discussion]{kon08, hol12}. Thus, before images could be interpreted based on thick-target emission of X-rays, high-resolution RHESSI spectra were analysed in order to separate nonthermal emission from thermal.

\subsection{Spectroscopy}

Spatially-integrated spectra were produced over the duration of the flare, using the same time intervals as those chosen for the imaging. Detector 4 on RHESSI was used due to its high spectral resolution of $\sim$1~keV at energies below $\sim$100~keV \citep{smi02}. Due to the low average count flux of the flare, significant noise was present, especially in the time before the HXR peak. This meant that, for many of the time intervals selected, various different combinations of thermal and nonthermal fit components could be used with equally good comparisons with data. These components included the thermal, thick target and Gaussian line options provided in the \textit{OSPEX} suite of algorithms \citep{kaa96}. It was found that the thermal component could be fit by a continuum variable thermal model, with a Gaussian line to account for the iron line complex emission at 6.7 keV. In some fit attempts, a thermal continuum component was not even necessary prior to the HXR peak. However, a full (line plus continuum) model could also be used to achieve equally good fits to the observed spectrum, based on the $\chi^2$ test provided in \textit{OSPEX}. In order to remain consistent with the thermal interpretation of the production of the iron line complex, the full thermal model was selected for this work.


During the fitting process, the sensivity of the fit to variation of the low-energy cutoff was investigated. This cutoff is a notoriously difficult parameter to derive from RHESSI spectra \citep[][]{sui05}, as only an upper limit to its value can usually be established. It was found that, for all time intervals prior to the HXR peak, the $\chi^{2}$ value of the fit was almost constant with different initial values of low-energy cutoff, ranging from 1 to 15~keV. This further indicates that the highest values that still produced good fits can only be seen as upper limits to this parameter. As such, the cutoff was assumed to be at an energy less than 5~keV for this analysis, which allowed the use of a smooth injection spectrum without a cutoff for the modelling. Further justification for this fully nonthermal interpretation is given by analysis of the images (see \S2.2).

As shown in Figure 1b, the resulting spectral fits show that, for the phase of the flare prior to the HXR peak, emission is predominantly nonthermal, except for that produced by the iron line at 6.7~keV. Given RHESSI's dynamic range of $\sim$ 1:10, it is likely that both types of emission are observable simultaneously below $\sim$7~keV \citep{hur02}.  It could be argued that this significant thermal emission is accounted for by the apparently thermal looptop source present during the early phase of the flare, even after the inital sources have descended down the loop. However, comparison of the total counts associated with this source and with the footpoint sources indicate that the looptop emission cannot alone produce all of the thermal emission indicated by the spectra. If the descending sources are produced by a injection of nonthermal electrons, there is expected to be localised heating and thus thermal emission at the site where energy deposition is at its peak. Therefore it may be the case that low-energy footpoint emission is a combination of thermal and nonthermal emission.

An estimate of the displacement between thermal and nonthermal footpoint emission can be made by approximating the distance covered by evaporating plasma over the time since the initial beam penetration. If ablation of chromospheric material begins at 04:35:48UT, and given standard evaporation velocities of $\sim$ 100--200~km s$^{-1}$, this would result in a displacement of thermal emission by $\sim$ 2.4--4.8 Mm at 04:36:12UT, the latest interval used in this study. In reality, the thermal source would be continually replenished by the ablation at the footpoint, so these displacement values are an upper limit only. Indeed, by modelling radiative and convective energy release following beam heating, \citet{all05} determine that, for an impulsive flare such as this, the displacement between the location of peak energy deposition and peak footpoint temperature can be as little as 0.3~Mm after 6~s for an impulsive event. Therefore, while the emission may have a thermal contribution, this desplacement error is small enough such that the X-ray sources of all energies will be used as a proxy for location of peak nonthermal energy deposition for the remainder of the paper. As such, it is appropriate to model their motion based purely on the location of the peak of the simulated nonthermal photon distribution with height, which is derived in \S 3.

\subsection{Imaging}

Images were reconstructed using the CLEAN algorithm, with detectors 3, 4, 5, 6, 8 and 9 \citep{hog74, hur02}. Detector 1 was excluded as the fine spatial resolution ($\sim$2.3~arcsec) tended to add small peak emission near the larger sources, making automated source-tracking unreliable. Detectors 2 and 7 were excluded as their imposed lower threshold energy is at least $\sim$9~keV.
 
Imaging revealed the motion of X-ray sources down and up the legs of the flare loop previously noted by \citet{sui06} (Figure 1c). A 3--10~keV source appears just west of the limb at 04:35:40~UT and descends $\sim$12~Mm down the apparent flare loop to reach the footpoints at 04:36:08~UT, which coincides with the peak in hard X-rays. Following this, the source rises $\sim$11~Mm to return to a looptop position, where it remains until soft X-ray emission returns to pre-flare level. This motion is seen in all three energy bands used for imaging, although the sources exhibit different qualitative behaviours before and after the HXR peak. Before the peak, the higher energy sources are located lower in the loop, descending slower and at different rates, covering $\sim$13~Mm $\sim$9~Mm and $\sim$3~Mm in the 3--6, 6--8 and 8--10~keV bands, respectively. After the peak, the distance travelled by each source is roughly constant with emitted photon energy, and higher energy emission originates higher in the loop, contrary to the ordering observed during the descent.

\begin{figure}
	\resizebox{\hsize}{!}{\includegraphics{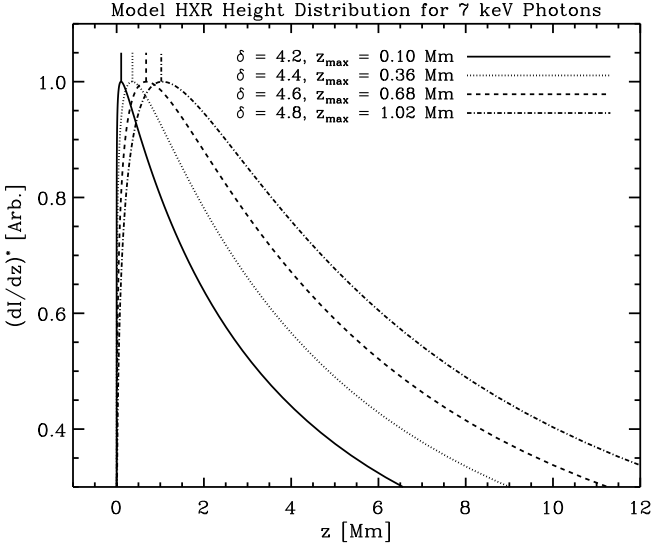}}
	\caption{Modelled HXR flux distribution with distance from the footpoint along the flare loop, produced using Equations (\ref{dIdzz}) and (\ref{Nz}), and assuming example density parameters $H_{r} = 10^{9} \mathrm{cm}$, $N_{r} = 2.43 \times 10^{19}\;\mathrm{cm^{-2}}$ and $a = 0.9$. Four sample spectral index values are input, at a photon energy of 7~keV. Since z- and $\delta$- independent factors are neglected, the distributions are normalised such that they peak at 1, however the location of the peaks and the relative scaling between plots of different index is accurate. The height at which dI/dz distributions are at their maximum ($z_{max}$) are noted, illustrating the HXR source height's dependence on spectral index. For lower, or harder, spectral indices, the height at which $dI/dz$ is at its maximum value is lower in the model flare loop.}
	\label{model}
\end{figure}

In order to compare with predictions of the thick target model, source position with time and energy was quantified (Figure 2). The southern leg of the loop was chosen for analysis, because the sources travelled further along this leg, resulting in better defined height values. The position of the source was represented by the peak of a 2-dimensional Gaussian fit to the southern CLEAN source, which was isolated by removing all flux lower than 30\% of the brightest pixel. Height was then defined as the distance from the southern footpoint to the position of the source along a curve that passed through these two points as well as the northern footpoint (Figure 2 inset).

The footpoints were defined as the peaks of the X-ray sources in the 20--50~keV range at the HXR peak of the flare (Figure 1c, orange contour). The height of these footpoints was used as a reference point for the heights of the low-energy sources. These relative heights were then converted to absolute heights by adding the predicted height of peak 25--50~keV emission, based on the CTTM (see \S 3). This analysis was repeated for all three energy bands used to create images.

With the evolution of the source height for each energy band quantified as a function of time, and values of nonthermal power-law index derived from spectra, the RHESSI observations were then compared directly to predicted height-time evolution based on the thick target model.

\begin{figure}
	\resizebox{\hsize}{!}{\includegraphics{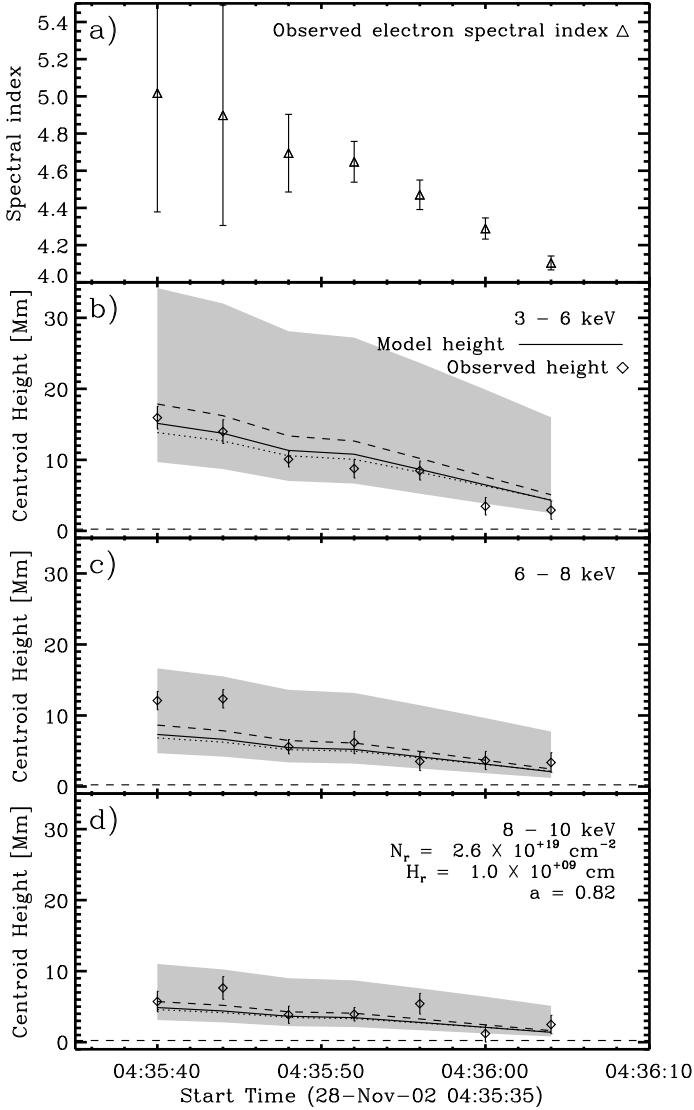}}
	\caption{(\textit{a}) Electron spectral index, based on fits to RHESSI spectra. (\textit{b--d}) Model and observed nonthermal source height evolutions for photon energies of 3--6~keV, 6--8~keV and 8--10~keV, respectively. Source heights derived from RHESSI observations are denoted by diamonds, with vertical solid lines indicating the 1-sigma width of the gaussian which was fit to the X-ray source. The heights corresponding to the peak in dI/dz are shown as a solid line. The shaded gray area extending above and below the solid line represents a `1-sigma' width of the model intensity distribution, calculated using its full-width half-max, where $FWHM = 2.35\sigma$. This serves to demonstrate the size and asymmetry of the predicted X-ray source. Two alternate model height evolutions are shown as dashed and dotted lines, which use different fit parameters $N_{r}$, $H_{r}$ and $a$. Along with the model given by the solid line, these all produce minimal $\chi^{2}$ values when fit to the data. The alternative results are presented to show the range of possible fits to the data, with corresponding density models shown in Figure 5. The dashed horizontal line represents the absolute height of the 25--50~keV footpoint, approximately 0.24~Mm. Fit parameters of the solid line are shown in section d. This model successfully accounts for a different apparent rate of descent for each emission energy, with sources observed at lower photon energy descending more rapidly.} 
	\label{htfit}
\end{figure}

\section{Thick target Modelling}

This section outlines the method by which a model nonthermal X-ray source height is calculated for a given injected spectral index, $\delta$ and photon energy, $\epsilon$. A power-law electron injection spectrum describes the distribution of electrons with their kinetic energy, $E_{0}$, before any interaction with coronal or chromospheric plasma, and has the form $f_{0}(E_{0}) = (\delta - 1)\;f_{1}/E_{1}\; (E_{0}/E_{1})^{-\delta}$, where $E_{1}$ and $f_{1}$ constitute a reference point in electron flux and energy. Following accleration, electrons travel down the flare loop and undergo Coulomb collisions with the ambient plasma, reducing their energy from $E_{0}$ to $E$. Thus, at a given distance, $z$, along the loop, the spectrum becomes $f(E, N(z))$, where $N(z) = -\int n(z) dz$ is the column depth and $n(z)$ is the number density of the ambient plasma \citep{bro72}. The energy lost to collisions is given by $E^{2} = E_{0}^{2} - 2KN$, \citep{bro72}, where $K=2\pi e^{4}\Lambda$ and $\Lambda$ is the Coulomb logarithm for an ionised plasma, which is used here as observed emission originates from heights at which the solar atmosphere is well-ionised \citep{bro75, ems78}.

In this work the goal is to determine the peak location of nonthermal X-ray emission by exploring different density models and injection spectral indices. \citet{bro02} derived this distribution of nonthermal photon flux with height as:
\begin{equation}
\frac{dI}{dz} = \frac{Af_{1}\sigma_{0}}{8\pi r^{2}E_{1}} (\delta - 1)\;\frac{1}{\epsilon}\; n(z) \; \left(\frac{E_{1}^{2}}{2KN(z)}\right)^{\delta/2} \; B\left(\frac{1}{1+u(z)},\frac{\delta}{2},\frac{1}{2}\right)
\label{dIdz2}
\end{equation}
where $r$ is the distance from source to observer, $A$ is the cross-sectional area of the loop, $u(z) = \epsilon^{2}/2KN(z)$ and $B(...)$ is the \textit{Incomplete Beta Function},
\begin{equation}
B\left(\frac{1}{1+u},\frac{\delta}{2},\frac{1}{2}\right) = \int^{1/(1+u)}_{0}{x^{\delta/2-1}\:(1-x)^{-1/2}\:dx}.
\end{equation}

As the peak position is the only parameter of interest for this work, for neatness we hereafter remove the constant factor $\alpha = Af_{1}\sigma_{0}/(8\pi r^{2}E_{r})$ and express the distribution as $(dI/dz)^{*} = (dI/dz)/\alpha$. From Equation (2) we therefore obtain:
\begin{equation}
\left(\frac{dI}{dz}\right)^{*} = (\delta - 1)\;\frac{1}{\epsilon}\; n(z) \; \left(\frac{E_{1}^{2}}{2KN(z)}\right)^{\delta/2} \; B\left(\frac{1}{1+u(z)},\frac{\delta}{2},\frac{1}{2}\right)
\label{dIdz2}
\end{equation}

In order to evaluate Equation \ref{dIdz2}, a model providing density \textit{n(z)} and column depth \textit{N(z)}, which are related for all z by $n(z)=-dN/dz$, is required. Using this relation one can say $n(z) = -N d(logN)/dz$ and define $H(N) \equiv -1/d(logN)/dz$ the \textit{local scale height}, such that \textit{n(z)} = \textit{N(z)/H(N(z))}. Thus a depth-varying scale height is implemented through the choice of \textit{H(N)}. In this work the chosen model for scale height is $H(N) = H_{r} (N_{r}/N)^{a}$, where $H_{r}$ and $N_{r}$ are reference scale heights and column depths, which along with \textit{a} can be varied freely, where $a>0$. It should be noted that, while this model is described by three variable parameters as presented, one can be set constant. As $H_{r}$ always appears in the factor $H_{r}N_{r}^{a}$, it will be left fixed at the constant value of $10^{9}$~cm, while $N_{r}$ and $a$ are allowed to vary. In order to constrain the model, limits can be set on \textit{n(z)} and \textit{H(N)} based on previously measured and physically expected values for the low solar atmosphere. 

Following this choice of \textit{H(N)}, Equation (\ref{dIdz2}) becomes
\begin{equation}
\left(\frac{dI}{dz}\right)^{*} = \frac{(\delta - 1)}{H_{r}N_{r}^{a}}\;\frac{1}{\epsilon}\;(E_{1}^{2}/2K)^{\delta/2} \; N^{1+a-\delta/2} B\left(\frac{1}{1+u},\frac{\delta}{2},\frac{1}{2}\right)
  \label{dIdzz}
\end{equation}
which can now be used to produce a plot of \textit{dI/dz} versus \textit{z} (see Figure 3), from which the height $z_{max}$ at peak \textit{dI/dz} can be calculated. In order to convert from a column depth to a height in the solar atmosphere, the relation $n(z) = -dN/dz = N/H(N) = N^{1+a}/(H_{r}N_{r}^{a})$ was used to form a differential equation, integration of which then gives

\begin{equation}
z_{max} = \frac{H_{r}}{a\left(N_{r}/N_{max}\right)^{a}}.
  \label{Nz}
\end{equation}
The limits of this integral are $N=N_{max}$ and $N=\infty$. As such this gives an absolute height of the model nonthermal source. For comparison with the observed source heights, which are measured as distance above the 25--50~keV footpoint, the model height of the footpoint is calculated and added to the observed values before comparison is made. It should be noted that this relatively small, roughly 0.25~Mm.

The model $(dI/dz)^{*}$ distribution for nonthermal emission of 7~keV photons is shown in Figure \ref{model}, at four different electron spectral indices. The position of peak emission is highlighted, illustrating the result that harder injection spectra (lower index values) result in a lower location of peak emission. This reflects the fact that if electrons are accelerated to higher kinetic energies, they will propegate further into the coronal plasma before losing their energy to long range Coulomb electron-electron interactions while radiating bremsstrahlung by short range electron-ion interactions.

This model will be used to produce expected nonthermal source heights based on the density model, spectral index and photon energy. These CTTM-based heights of peak photon flux can be fit to those recorded by RHESSI, using the electron spectral index evolution, $\delta(t)$ provided by the fits to RHESSI spectra. A close match between the observed and modelled source heights will indicate whether or not the CTTM prediction of source descent is a possible interpretation of the observed motions in this flare. Additionally, the density model required by the fit can be compared with previous observations given in other work in order to determine if the densities required to produce this result are commonly encountered in flaring plasma.

\section{Results}

\begin{figure}[ht!]

	\resizebox{\hsize}{!}{\includegraphics{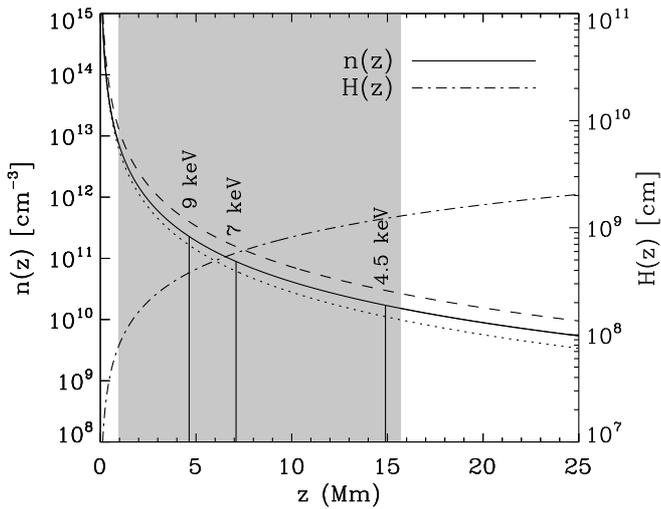}}
	\caption{ Density profile required to obtain the fits shown in Figure 4. Density ($n(z)$) and local scale height ($H(z)$) versus height, $z$, above the footpoint are shown. The input parameters $H_{r}$, $N_{r}$ and $a$ are the best-fit parameters resulting from the fitting process outlined in Figure 4. Two alternate density models, which are derived from the alternate fits shown in Figure 4, are given here as dashed and dotted lines. The shaded region represents the range of heights within which observations of HXR sources were made, and so densities and scale heights outside of these range are not expected to be accurate. It was assumed that the density structure of the flare plasma was approximately constant over the $\sim$20~s rise phase of the HXRs. Vertical solid lines indicate the location of peak emission for the denoted energies, which represent the three energy bands used in this study. The values shown correspond to the first observation, where the injection spectral index is $\delta = 5$.
}

	\label{denmod}
\end{figure}

The comparison between model and observed height-time evolution is shown in Figure \ref{htfit}. While heights were determined earlier and later in the flare, only the portion of the height-time evolution that was fitted with our model is shown. The first time interval was not used as images and spectra were noise-dominated, and so both the measured height value and spectral index were inaccurate. Data after 04:36:00UT have also been neglected from the fitting algorithm, as it is believed here that the emission is becoming predominantly thermal at 3--10~keV, and so is not expected to be predictable based on a nonthermal electron flux model. Vertical lines at each data point represent one-sigma widths of the Gaussians that were fit to the RHESSI sources, which remained around roughly 2--3~Mm, corresponding to the RHESSI psf's HWHM.

An initial observation of importance is the distribution in height for the three energies before the HXR peak at 04:36:12 UT. The 8--10~keV source is located lower in the loop than the 6--8~keV source, which likewise is lower than the 3--6~keV source. This distribution does not hold for the full duration of the flare; there is a reversal at the HXR peak of the flare (Figure \ref{heighttime}). In the nonthermal scenario, a flux of nonthermal electrons travels through an increasingly dense chromospheric plasma. Higher-energy electrons are stopped by higher densities, and so the fastest electrons will penetrate deeper before their bremsstrahlung emission peaks. So, in this regime, high-energy emission is expected to be located lower in the loop than low-energy emission. However, for thermal emission, the reverse is true if magnetic reconnection is progressing above the loop. In this scenario, the upper loops are newly-reconnected and hotter, while plasma underneath has had time to cool, leading to the highest energy thermal emission being located nearer the looptop \citep{tsu92}. Keeping this in mind, the imaging analysis performed here suggests that the tracked X-ray emission in the 3 - 10~keV energy band is nonthermal until the HXR peak, at which point thermal emission becomes dominant as the sources appear to rise. This is consistent with the spectroscopic results, and further suggests the the tracked emission before the HXR peak can be treated as nonthermal, and so it is during this phase that the CTTM can be fitted to the data.

As shown in Figure \ref{htfit} b--c, for suitably chosen model scale height parameters, a model source descent can be simulated which shows strong agreement with observation. Three source descents are shown, one for each of the three energy bands used for imaging. The 3--6~keV source first appears $\sim$15~Mm above the footpoint before descending to a height of $\sim$5~Mm, while the 6--8 and 8--10~keV emission appears at $\sim$12 and $\sim$5~Mm respectively, all reaching approximately the same height above the footpoint. This difference in apparent rate of descent of nonthermal source was predicted in the CTTM through the use of a depth-varying hydrostatic scale height, an essential part of the density model used in the fitting process.

This density model is summarised in Figure 5. In order to help constrain the fit parameters, it was ensured that the resulting density and scale height models agreed reasonably with previous observations \citep[e.g.,][]{liu06,asc02}.

\section{Discussion and Conclusion}

In order to treat the observed source motion with the CTTM, energies at which emission could be considered nonthermal first needed to be established. Spectroscopic analysis suggests that, prior to the HXR peak, emission is predominantly nonthermal above 7~keV, and contributed to by both thermal and nonthermal components below that energy. As the flare progresses, the lowest energy bands become dominated by thermal emission. This can be explained by heating of the plasma in the flare loop from 8~MK to 11~MK within 15~s, as derived from GOES observations, using the background subtraction method outlined in \citet{rya12}. As the plasma reaches greater temperatures, it emits thermal radiation at higher energies. This heating period is expected to take place in all flares, however in this case it was gradual enough to allow a significant amount of low-energy HXR detections to be made. Therefore, it was deemed appropriate to analyse observed source motions based on the CTTM.

A close match between model and observed X-ray source heights were obtained in this work (Figure 4), however many important assumptions were made in order to do so, including that of a model density structure. Densities ranging from $10^{11}$ to $10^{13}~$cm$^{-3}$ over 20~Mm within a flare loop have been observed in previous RHESSI studies \citep[e.g.,][]{liu06}. The required density distribution in this work show similar structure, and are also in line with derived densities of \citet{asc02}. An interesting requirement for this analysis was the introduction of a depth-varying scale height, which is responsible for the difference in apparent descent rate of the nonthermal emission between different photon energies. Close to $z=0$, the required scale height is on the order of $10^{7}$~cm, or hundreds of kilometers, in agreement with previous RHESSI-based calculations of $\sim$130--140~km \citep{kon08b,sai10}, as well as with scale heights derived from temperatures put forward by modelling of visible and UV emission \citep{ver81}. The latter work as well as that laid out by \citet{all05} also suggest coronal temperatures consistent with a scale height on the order of a number of megametres, as was also required by this fit. Without a variation in scale height (and thus, temperature), the distance between sources of different energy would be constant, contrary to observations of this event.

It was shown in the process of modelling the distribution of nonthermal emission with height that a strong asymmetry should be present in observed sources. As CLEAN was used to reproduce the images, it may be the case that this asymmetry was diminished, but also that the peak of the model distribution could be shifted by the reconstruction process. To test this, the model intensity distributions were run through a one-dimensional version of CLEAN. For a small number of iterations (100), some loss of the asymmetry of the source was seen, with the resulting distribution approaching a gaussian shape, which may explain the near-gaussian shape of the sources in RHESSI images. The resulting peak position was seen to shift by, at most, $\sim$1~Mm in the case of the model used. Finally, the presence of a low energy cutoff substantially higher than the photon energy would have the effect of removing the low-energy electrons that contribute most to the 'tail' of the asymmetric model source. In this way, a cutoff could also explain the observation of symmetric sources.

The model used in this work relied on the assumption that the thick target model is accurate, and that the density structure of the target is the dominating factor on X-ray source position. It should be noted that other relevant mechanisms have been discussed but were not taken into account here. One could consider pitch-angle diffusion, where immediately following energy release, electron flux exhibits a large pitch angle, and so is contained to the higher parts of the loop \citep{fle97}. Over time, diffusion causes a lowering in pitch angle and the bulk of the accelerated electrons move gradually further down the loop, which may contribute to a source descent. Another important consideration concerns the evolution of $n(z)$ with time. As electrons are accelerated into the flare loop, they cause heating and expansion which results in a redistribution of local plasma density, which should lead to a prediction of a rise of nonthermal HXR sources. This would work against mechanisms which cause a descent in HXR emission. \citet{bat12} make use of Fokker-Planck modelling to determine the degree by which various mechanisms displace peak heights from their location as determined by collisional effects alone. They find that overall, displacements of $\sim$10\% in source position can be caused by magnetic mirroring and the implementation of a non-uniformly ionised flare loop, while pitch-angle scattering can cause a more stark displacement of up to 20\%. It would therefore be important to allow for these effects in a complete model.

Keeping these remarks in mind, it has been shown here that the hardening of the electron injection spectrum is, with suitably chosen model densities and injection spectrum, sufficient to drive downward motion of nonthermal X-ray sources during the initial stage of SHS evolution. This model requires that there is indeed a flux of electrons from the looptop, or at least from $\sim$20~Mm above the footpoint towards the footpoints of the flare loop. Models invoking torsional Alfv\'en waves as the mechanism of primary energy transfer from the corona \citep{fle02} or cascading reconnection in the chromosphere with re-acceleration there \citep{bro09} as yet offer no explanation of such a relation between spectral index and HXR source heights. However, following further developement of these relatively new models, events such as this may be useful in testing these predictions of nonthermal source behaviour before the peak in HXRs.

\begin{acknowledgements}

The author would like to thank Brian Dennis and the RHESSI team for their hospitality and input during this research. This work has been supported by a Government of Ireland Studentship (AMO'F) from the Irish Research Council for Science, Engineering and Technology (IRCSET) and a Visiting Fellowship by TCD (JCB). GDH acknowledges support from the NASA Heliophysics Guest Investigator Program and the RHESSI project.

\end{acknowledgements}

\bibliographystyle{aa}
\bibliography{2012AAdraft} 

\end{document}